\documentclass[aps,twocolumn,preprintnumbers,amsmath,amssymb,superscriptaddress,prl,10pt]{revtex4-1}
\usepackage{amsfonts}
\usepackage{amsmath}
\usepackage{amssymb}
\usepackage{graphicx}
\usepackage{textcomp}
\usepackage{siunitx}

\newcommand{\mytitle}{Second-Harmonic Generation in Silicon Nitride Ring Resonators}

\newcommand{\xtwo}{$\chi^{(2)}$}
\newcommand{\xthree}{$\chi^{(3)}$}
\newcommand{\nitride}{Si$_3$N$_4$}
\newcommand{\oxide}{SiO$_2$}
\newcommand{\micron}{$\mu$m}

\begin{document}

\title{\mytitle}
%\author{Jacob~S.~Levy$^{1}$, Mark~A.~Foster$^{2}$, Alexander~L.~Gaeta$^{2}$, Michal~Lipson$^{1,3}$} 
%\email{jsl77@cornell.edu}
%\homepage{http://nanophotonics.ece.cornell.edu}
%\affiliation{$^{1}$School of Electrical and Computer Engineering, Cornell University, Ithaca, NY 14853}
%\affiliation{$^{2}$School of Applied and Engineering Physics, Cornell University, Ithaca, NY 14853}
%\affiliation{$^{3}$Kavli Institue at Cornell for Nanoscale Science, Cornell University, Ithaca, NY 14853}
\author{Jacob S. Levy}
\email{jsl77@cornell.edu}
\affiliation{School of Electrical and Computer Engineering, Cornell University, Ithaca, NY 14853}

\author{Mark A. Foster}
\affiliation{School of Applied and Engineering Physics, Cornell University, Ithaca, NY 14853}
\author{Alexander L. Gaeta}
\affiliation{School of Applied and Engineering Physics, Cornell University, Ithaca, NY 14853}
\author{Michal Lipson}
\homepage{http://nanophotonics.ece.cornell.edu}
\affiliation{School of Electrical and Computer Engineering, Cornell University, Ithaca, NY 14853}
\affiliation{Kavli Institue at Cornell for Nanoscale Science, Cornell University, Ithaca, NY 14853}

\begin{abstract}
The emerging field of silicon photonics seeks to unify the high bandwidth of optical communications with CMOS microelectronic circuits.  Many components have been demonstrated for on-chip optical communications, including those that utilize the nonlinear optical properties of silicon\citep{Foster06,Corcoran09}, silicon dioxide\citep{Kipp04,Carmon07} and silicon nitride\citep{Ikeda08,Levy10}.  Processes such as second harmonic  generation, which are enabled by the second-order susceptibility, have not been developed since the bulk \xtwo{} vanishes in these centrosymmetric CMOS materials.  Generating the lowest-order nonlinearity would open the window to a new array of CMOS-compatible optical devices capable of nonlinear functionalities not achievable with the \xthree{} response such as electro-optic modulation, sum frequency up-conversion, and difference frequency generation.  Here we demonstrate second harmonic (SH) generation in CMOS compatible integrated silicon nitride (\nitride{}) waveguides.  The \xtwo{} response is induced in the centrosymmetric material by using the nanoscale structure to break the bulk symmetry.  We use a high quality factor \textit{Q} ring resonator cavity to enhance the efficiency of the nonlinear optical process and detect SH output with milliwatt input powers.
\end{abstract}

\maketitle

Silicon nitride is a centrosymmetric, CMOS compatible material shown to be useful for integrated optics and to possess interesting nonlinear properties\citep{Ikeda08,Levy10,Tan10}. Previous nonlinear experiments in silicon nitride waveguides have only used the \xthree{} nonlinearity: the lowest order in the material since the bulk electric dipole response vanishes. In this study we induce a \xtwo{} response utilizing the interface between two centrosymmetric materials, the \nitride{} core and the silicon dioxide (\oxide{}) cladding.  The waveguide interface breaks the bulk symmetry and a second-order nonlinear response can arise\citep{Bloembergen68,Tom83,Sipe87,Shen89} from the asymmetric dipole potential formed at the surfaces.  Previously, detecting the second harmonic wave from a reflected interface has been used for monitoring surface properties\citep{Tom83,Shen89}, even with centrosymmetric materials, however no CMOS-compatible integrated devices have shown guided SH. 

\begin{figure}[!htp]
\begin{center}
\includegraphics[width=1\linewidth]{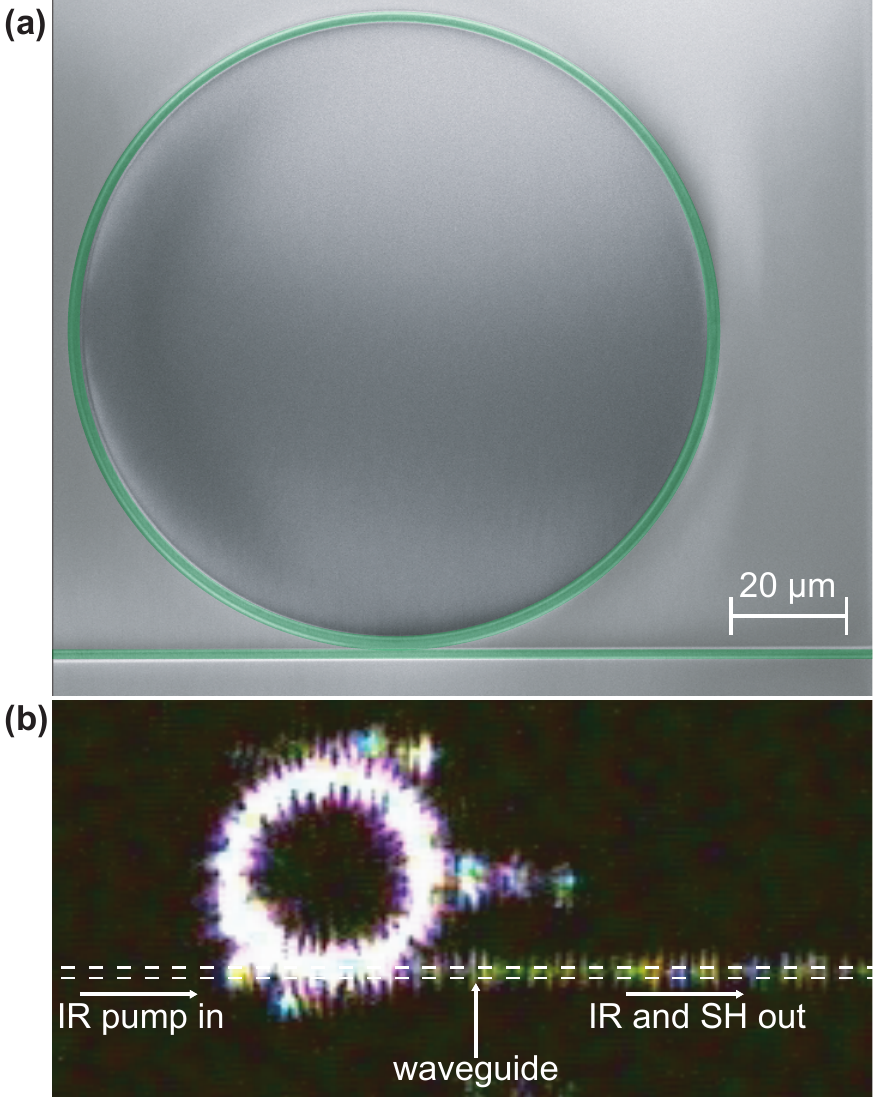}
\end{center}
\caption{\textbf{Silicon nitride ring resonator for SH generation.}
\textbf{a.} A scanning electron micrograph image of a typical silicon-nitride ring microresonator. 
\textbf{b.} Top view visible CCD camera image of the microresonator generating red light.  IR light, invisible to this camera, is launched from the left and couples into the ring.  The power builds-up in the ring generating SH which couples back into the waveguide propagating out to the right.}
\label{fig1}
\end{figure}

\begin{figure*}[bhtp]
\begin{center}
\includegraphics[width=1\textwidth]{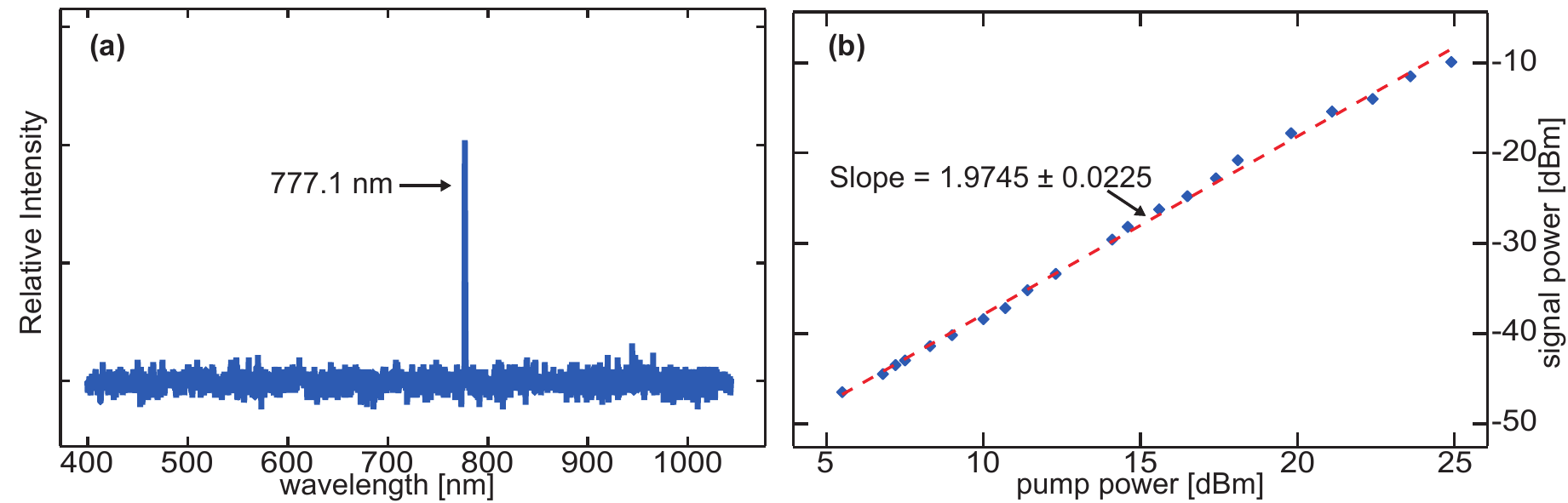}
\end{center}
\caption{\textbf{The spectral and power output from the generated SH light.}
\textbf{a.} The output from the waveguide, pumped at \SI{1554.2}{nm}, is collected into an optical spectrometer.  The visible emitted light occurs only at the expected SH wavelength.
\textbf{b.} The power dependence of generated SH versus pump power dropped to the ring.  The red-dashed line with slope near 2 on the log-log plot represents the best fit line to the data and is very close to the theoretical square prediction.}
\label{fig2}
\end{figure*}

We observe SH generation in this experiment using a device based on recently demonstrated \nitride{} ring resonators with both high \textit{Q} and high modal confinement\citep{Gondarenko09}.  Integrated ring resonators enhance the efficiency of nonlinear interactions\citep{Levy10,Turner08,Ferrera08} due to the large power build-up in the ring.  We design silicon nitride ring resonators to be resonant at both fundamental and SH frequencies with the same fabrication process previously described\citep{Levy10,Gondarenko09}.  The ring resonator (Fig.~\ref{fig1}a) used in this experiment has a radius of 116 \micron{} and a coupling gap of \SI{350}{nm} between the ring and the bus waveguide.

We measure the SH wavelength in the waveguide coupled to the resonator for input pump powers as low as \SI{3}{mW}.  A tunable diode laser is amplified using an erbium-doped fiber amplifier and coupled to the waveguide using a tapered lensed fiber.  We use a polarization controller to launch light in the fundamental quasi-TE mode of the waveguide.  The pump wavelength~$\lambda_{P}$ is tuned into the resonance of the ring cavity near \SI{1554}{nm} and, with suitable power levels, we are able to observe generation of the second harmonic (Fig.~\ref{fig1}b).  We use a spectrometer to measure the wavelength of the visible emitted light. Figure~\ref{fig2}a displays the output with the generated SH wavelength~$\lambda_{SH}$ measured to be \SI{777.1}{nm} which, as expected, is~$\lambda_{P}/2$.  Since only the SH wavelength is detected, we conclude we are not generating broadband photoluminescence which has been previously described in silicon nitride\citep{Barth08,Serpenguzel01}.  To measure the power of the SH, both the IR pump and the visible signal are collected into an optical spectrum analyzer (OSA).  The powers measured by the OSA are corrected to the absolute power values coming out of the waveguide, taking into consideration the coupling efficiencies from the waveguide to the fiber and to the OSA for both pump and SH wavelengths, respectively.  The OSA confirms the wavelength measurement for both the pump and SH.  We observe a maximum conversion efficiency of -35 dB with \SI{100}{\micro\watt} of SH generated for a pump of \SI{315}{mW}.  At increased pump powers, the ring's resonance experiences a severe thermal shift which prevents efficient coupling from the waveguide to the resonator.  In order to clearly demonstrate the theoretical square dependence of the SH process, we plot the dropped pump power against the generated SH on a log-log scale (Fig.~\ref{fig2}b) and calculate a best fit slope of $1.9745 \pm 0.0225$.

\begin{figure*}[!htb]
\begin{center}
\includegraphics[width=1\textwidth]{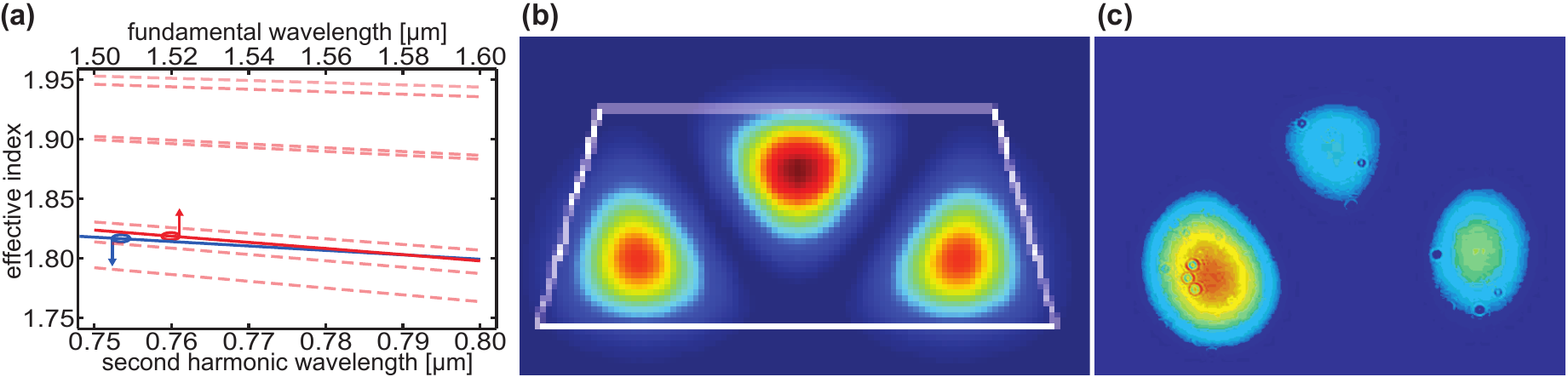}
\end{center}
\caption{\textbf{Calculation of mode-order for phase matched SH generation.}
\textbf{a.} The solid blue line represents the effective index of the fundamental quasi-TE mode at the IR pump as a function of wavelength. The dashed red lines represent the effective indices for modes of the corresponding SH wavelengths.  The sixth-order mode of the SH (solid red) has a crossing with the blue line indicating a point of phase-matching. 
\textbf{b.} The simulated cross-section mode profile for the sixth-order mode of our waveguide at 777.1 nm.
\textbf{c.} The captured mode image of the visible emission from our waveguide showing good agreement with the simulated mode profile.}
\label{fig3}
\end{figure*}

Efficient SHG is possible by satisfying the phase-matching condition of matching the effective index of the fundamental waveguide mode in the IR to a higher order mode in the visible range. This method has previously been suggested for phase-matched third-harmonic generation in \oxide{}, in both microtoroids\citep{Carmon07} and microstructured fibers\citep{Omenetto01,Ranka00}. In order to determine which mode is best phase-matched, we use a finite difference method mode solver to calculate the effective index for both the pump and SH frequencies.  Figure~\ref{fig3}a, shows a plot of the effective index against wavelength for the first $8$ modes of the waveguide at the SH wavelength and the fundamental mode at the pump wavelength.  Since the phase-matching condition is satisfied when the effective index at the pump and corresponding SH wavelengths are equal, crossing points between the blue and red lines are perfectly phase-matched for the corresponding wavelength pair and mode number.  As the plot shows, we have a phase-matching point near our experimental pump wavelength with the $6^{\mathrm{th}}$ order mode of the SH.

The captured mode image of the waveguide output (Fig~\ref{fig3}c) corresponds well to the simulated mode profile for the $6^{\mathrm{th}}$ order SH mode which optimizes the phase-matching condition.  Since the ring and bus waveguide cross-sections are identical, we expect the same-order mode generated in the ring resonator to couple to the waveguide.  To effectively image the mode, we polish away the nanotapers\citep{Almeida03}, used to increase coupling efficiency, at the output of the waveguide.  We collect the output light with a high NA objective to focus the image on a CCD camera.  There are three distinct lobes in the mode showing a good match with the simulated mode.

We calculate the effective \xtwo{} to be \SI{4d-14}{\metre\per\volt} from the conversion efficiency observed in the ring.  For SHG with an undepleted pump, the expected signal power may be calculated for a given pump intensity and propagation distance by solving the coupled amplitude equations\citep{Boyd08}.  Since we are using a resonator, the intensity of the pump and SH are increased by the respective cavity enhancement effects of the ring.  In order to accurately model the nonlinear susceptibility, we take into account the finesse of the cavity, the simulated modal field overlap and phase-mismatch, and the radius of the ring.  The power $P_{sh}$ for second harmonic wave is then given by: 

\begin{equation}
P_{sh}=\frac{C_{sh}C^2_p(\omega_p\chi^{(2)}LP_p)^2}{8n^2_p n_{sh}c^3\epsilon_0A^2_p}A_{sh}\mathrm{sinc}^2(\Delta kL/2)f(A_p,A_{sh})
\label{eq:1}
\end{equation}

\begin{equation}
C_{i} = \frac{P_{circ}}{P_{in}} = \left|\frac{j\kappa_{i}\mathrm{exp}(-\alpha_{i}L/2)}{\mathrm{exp}(jk_{i}L)-\tau_{i}\mathrm{exp}(-\alpha_{i}L/2)}\right|^{2}
\label{eq:2}
\end{equation}

where $\omega_p$ is the pump frequency, $\epsilon_0$ is the permittivity of free space, $c$ is the speed of light in vacuum, $n_i$ is the effective index for the modes, $A_i$ is the mode area, $\Delta k$ is the phase mismatch, $L$ is the ring circumference and $P_p$ is the pump power in the waveguide.  The modal overlap integral between the fundamental and second-harmonic fields is accounted for by the function $f(A_p,A_{sh})$.  $C_i$ takes into account the circulating power in the ring\citep{Turner08} where $\kappa$ and $\tau$ represent the coupling parameters from the waveguide to the ring, $\alpha$ is the propagation loss in the ring and $k_i$ is the wavenumber.  From Eq.~\ref{eq:1}, we see that the SH wave has a quadratic dependence on the pump power and nonlinear susceptibility \xtwo{}.  From the measured output power values we estimate our induced \xtwo{}.  The integrated high finesse resonator in silicon nitride increases the efficiency of the SHG significantly when the pump and SH are both resonant.

In addition to SH, we also observe third harmonic generation in similar ring resonators generating green light with a wavelength of \SI{520}{n\metre}.  The \xthree{} nonlinearity of the silicon nitride induces the third order process in which the fundamental mode is phase-matched to the $18^{\mathrm{th}}$ order mode at the third harmonic wavelength.  
We collect \si{\pico\watt}'s of light at the third harmonic wavelength from a pump centered at a cavity resonance near \SI{1560}{nm}.  This is on the same order as shown in a silicon photonic crystal\citep{Corcoran09}, but here light is coupled and guided in the bus waveguide as opposed to out of plane emission.

Our demonstration of guided on-chip visible light generation opens the available spectrum for Si-based devices from the IR to the visible, increasing bandwidth and enabling potential integration of silicon photodetectors to on-chip optical networks.  Additionally, the doubly resonant SH generation presented here could produce squeezed states\citep{Pereira88} of both the pump and SH frequencies for quantum optics studies.  Finally, the induced second-order nonlinearity could be used for difference frequency generation to combine two near infrared pumps to generate a mid-infrared source\citep{Hon09}.

\section*{acknowledgements}
The authors would like to acknowledge DARPA for supporting this work under the Optical Arbitrary Waveform Generation Program and the MTO POPS Program and by the Center for Nanoscale Systems, supported by the National Science Foundation and the New York State Office of Science, Technology and Academic Research.   This work was performed in part at the Cornell NanoScale Facility, a member of the National Nanotechnology Infrastructure Network, which is supported by the National Science Foundation (Grant ECS-0335765).

\bibliographystyle{naturemag}
\bibliography{SHG}

\end{document}